\documentclass[a4paper,twoside]{article}
\newcommand{\affil}[1]{$^{\rm #1}$}
\usepackage[authoryear]{natbib}
\bibpunct{(}{)}{;}{a}{}{,}
\usepackage{graphicx}
\date{} 
\title{\large\bf\flushleft New candidate Planetary Nebulae in the IPHAS survey: the case of PNe with ISM interaction.}
\author{\parbox{\textwidth}{\flushleft
\vspace{-0.5cm}
{\it Laurence Sabin\affil{A}, Albert A. Zijlstra\affil{A}, Christopher Wareing\affil{B}, Romano L.M. Corradi\affil{C}, Antonio Mampaso\affil{C}, Kerttu Viironen\affil{C}, Nicholas J. Wright\affil{D} and Quentin A. Parker\affil{E}}\\
\vspace{0.4cm}
{\small \affil{A}\,Jodrell Bank Center for Astrophysics, School of Physics and Astronomy, University of Manchester, Manchester M13 9PL, UK}\\
{\small \affil{B}\,Department of Applied Mathematics, University of Leeds, Leeds, LS2 9JT, UK}\\
{\small \affil{C}\,Instituto de Astrofisica de Canarias, Tenerife, Spain}\\
{\small \affil{D}\,Harvard-Smithsonian Center for Astrophysics, 60 Garden Street, Cambridge, MA, 02138, USA}\\ 
{\small \affil{E}\,Macquarie University/Anglo-Australian Observatory, Department of Physics, North Ryde, Sydney NSW 2190, AUSTRALIA}\\ 
{\small \affil{A}\,Email: laurence.sabin@manchester.ac.uk}}}
\begin{document}
\twocolumn[
\begin{changemargin}{.8cm}{.5cm}
\begin{minipage}{.9\textwidth}
\vspace{-1cm}
\maketitle
%
%
\small{\bf Abstract:}  We present the results of the search for candidate Planetary Nebulae interacting with the interstellar medium (PN-ISM) in the framework of the INT Photometric H$\alpha$ Survey (IPHAS) and located in the right ascension range 18h-20h. The detection capability of this new Northern survey, in terms of depth and imaging resolution, has allowed us to overcome the detection problem generally associated to the low surface brightness inherent to PNe-ISM. We discuss the detection of 21 IPHAS PN-ISM candidates. Thus, different stages of interaction were observed, implying various morphologies i.e. from the unaffected to totally disrupted shapes. The majority of the sources belong to the so-called WZO2 stage which main characteristic is a brightening of the nebula's shell in the direction of motion. The new findings are encouraging as they would be a first step into the reduction of the scarcity of observational data and they would provide new insights into the physical processes occurring in the rather evolved PNe.\\


\medskip
\medskip
\end{minipage}
\end{changemargin}
]
\small

\section{Introduction}

Large H$\alpha$ surveys have so far allowed the detection of $\sim$ 3000
planetary nebulae (PNe) in the Galaxy. The data can be principally found in
the Strasbourg-ESO Catalogue \citep{Acker} and the recent
Macquarie-AAO-Strasbourg H$\alpha$ Planetary Nebula Catalogues: MASH I and II
(\citet{parker2006} and \citet{Miszalski2008}). Unfortunately a limitation in
our understanding of this short and rather complex phase of stellar evolution
lies either in the deepness of the detections realised or the type of PNe
investigated. Indeed, although enormous progress has been made over the years
in terms of observations, the well-studied PNe  are generally bright and
often young. This hampers the study of:\\
\begin{itemize}
\item PNe hidden by the interstellar medium, particularly those located at low galactic height.\\
\item PNe with (very) low surface brightness where we find the group of old PNe.\\
\item Very distant PNe which appear as unresolved and not recognisable as nebulae.\\
\item PNe located in crowded areas such as the galactic plane. \\
\end{itemize}
Moreover, excluding these objects from global studies (morphology, abundances,
luminosity function...etc) may bias our understanding of planetary nebulae. As
an illustration, few PNe are described in the literature as ``PNe with ISM
interaction'', which is the step before the complete dilution of the nebulae
in the interstellar medium (\citet{Borkowski1990}, \citet{Ali2000},
\citet{Xilouris1996} and \citet{Tweedy1996}). The study of the interaction
process would give new insights into several aspects of the PN
evolution. Indeed, the density difference between ISM and PNe will affect
their shape. This is expected to be observable in old objects where the nebular
density declines sufficiently to be overcome by the ISM density. Other
phenomena like the flux and brightness enhancement following the compression
of the external shell, the increase of the recombination rate in the PN
\citet{Rauch2000}, the occurrence of turbulent Rayleigh-Taylor
instabilities and the implication of magnetic fields \citet{Dgani1998} are
among the physical processes which need to be addressed not only from a
theoretical but also observational point of view.  

The low surface brightness generally associated to PNe-ISM has for a long time
prevented any deeper observation and good statistical study of these
interactions, where only the interacting rim is well seen. New generations of
H$\alpha$ surveys have overcome this problem. A perfect example is the
discovery of PFP 1 by \citet{Pierce2004} in the framework of the AAO/UKST
SuperCOSMOS H$\alpha$ survey (SHS) \citep{Parker2005}. This PN, starting to
interact with the ISM at the rim, is very large (radius = 1.5 $\pm$0.6 pc) and
very faint (logarithm of the H$\alpha$ surface brightness equal to -6.05
erg\,cm$^{-2}$.s$^{-1}$.sr$^{-1}$).  In order to unveil and study
this ``missing PN population'' in the Northern hemisphere we need surveys
providing the necessary observing depth: the Isaac Newton Telescope (INT)
Photometric H$\alpha$ Survey (IPHAS) is one of them and will complete the work
done in the South by the SHS.\\

\section{IPHAS contribution}
IPHAS is a new fully photometric CCD survey of the Northern Galactic Plane,
started in 2003 (\citet{DREW2005}, \citet{Solares08}) and which has now been
completed \footnote{http://www.iphas.org}. Using the 2.5m Isaac Newton
Telescope (INT) in La Palma (Canary Islands, SPAIN) and the Wide Field Camera
(WFC) offering a field of view of 34.2$\times$34.2 arcmin$^{2}$, IPHAS targets
the Galactic plane in the Northern hemisphere, at a latitude range of
-5$^\circ$ $<$ b $<$ 5 $^\circ$ and covers 1800 deg$^2$. This international
survey is conducted not only in H$\alpha$ but also makes use of two continuum
filters, respectively the Sloan r' and i'. IPHAS is viewed as an enhancement
to former narrow-band surveys, first due to the use of CCD and the
particularly small pixel scale allowed by the WFC with 0.33 arcsec pix$^{-1}$
but also (and mainly) due to the depth reached for point sources
detection. Thus sources with a r' magnitude between 13 and 19.5-20 could be
detected with a very good photometric accuracy. The most interesting
characteristic for our purpose is the ability to detect resolved extended
emissions with an H$\alpha$ surface brightness down to 2$\times$10$^{-17}$ erg
cm$^{-2}$ s$^{-1}$ arcsec$^{-2}$ .\\

\begin{figure}[!h]
\begin{flushleft}
\includegraphics[height=7cm]{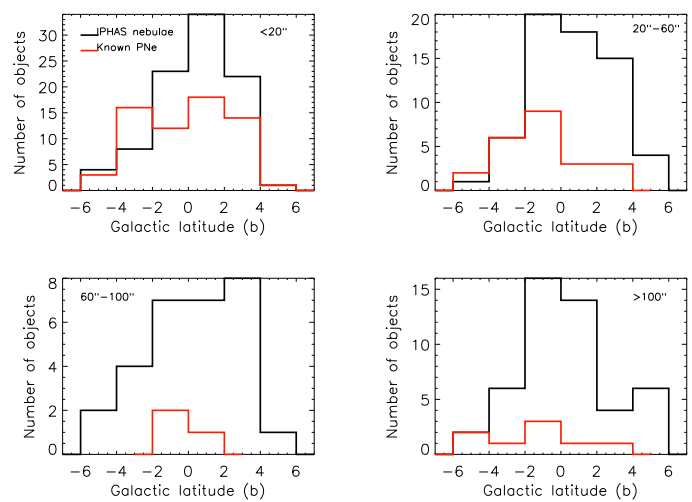}
\caption{Galactic distribution of the IPHAS nebulae according to their size.}\label{size}
\end{flushleft}
\end{figure}

In this paper we will focus on extended (candidate) PNe (i.e. objects with a
size greater than 5 arcsec). They were searched for via a visual inspection of
2 deg$^2$ H$\alpha$-$r$ (continuum removal) mosaics made from the different
IPHAS observations. And in order to allow the detection of objects of multiple
size and brightness level, the mosaics were binned at respectively 15
pixels$\times$15 pixels (5 arcsec) and 5 pixels$\times$5 pixels (1.7
arcsec). The first binning level, which is of particular interest to us, helps
to detect resolved, low surface brightness objects (down to the IPHAS limit)
and to accentuate the contours/shape of the nebulae (this is particularly
useful to see, for example, the full extent of an outflow or a tail). The
second set, is used to detect intermediate size nebulae i.e. smaller than
$\sim$15-20 arcsec in diameter.\\ The first area that has been fully
investigated is the region between RA=18h and RA=20h. We detected 233
candidate PNe among which other nebulosities may be found e.g. small HII
regions (Sabin, PhD thesis, to be published). Around 20$\%$ of this sample have
been so far spectroscopically confirmed as PNe (Sabin et al., in preparation).
If we look at the particularities of the PNe and candidate PNe uncovered, we
observe that from the point of view of the size, large objects (greater than
20 arcsec) constitute the main new group (Fig. \ref{size}). As large objects
are generally considered as more evolved, we are confident in finding in this
group new old PNe and by extension new cases of PNe interacting with the
surrounding ISM (PNe/ISM).

\section{Candidate PNe with ISM interaction}

Fundamental in PN development, the interaction with the ISM does not only
concern old PNe, as may be commonly thought. Indeed, the PN-ISM interaction
has mainly been detected in a rather small number of nebulae, which are
generally bright objects (``young'' and ``mid-age'' PNe). 
\citet{Rauch2000} and \citet{Wareing07} showed that different stages of
interaction are exhibited during the PNe life. The low surface brightness,
generally associated with nebulae mixing with the ISM and ``old'' PNe, has for
a long time prevented any deeper observation and good statistical study of
these interactions. Although faint objects will still remain difficult to
detect, the IPHAS survey provides a noticeable improvement. Nevertheless, a
caveat is the difficulty to visually separate PNe-ISM from other faint and
extended structures like old HII regions, Supernovae (SNRs) or diffuse
H$\alpha$ structures. As an example, faint bow shocks generally
characteristics of PNe mixing with the ISM can also be filamentary structures
from old SNRs. A spectroscopic analysis is the only way to have a clear
identification.\\ The work presented here is based on the classification from
\citet{Wareing07} (WZO 1-4 called after the authors' names) and will allow us
to establish the degree of interaction for each nebula. Their classification
is the result of the first extensive investigation of the applicable parameter
space, varying stellar parameters, relative velocities through the ISM and ISM
densities.\\ The depth reached by the IPHAS survey combined with the binning
detection method allowed us to identify 21 cases of interacting candidate
PNe. \\

\subsection{WZO1 type}

\begin{figure*}[h]
\begin{center}
\includegraphics[scale=0.36, angle=0]{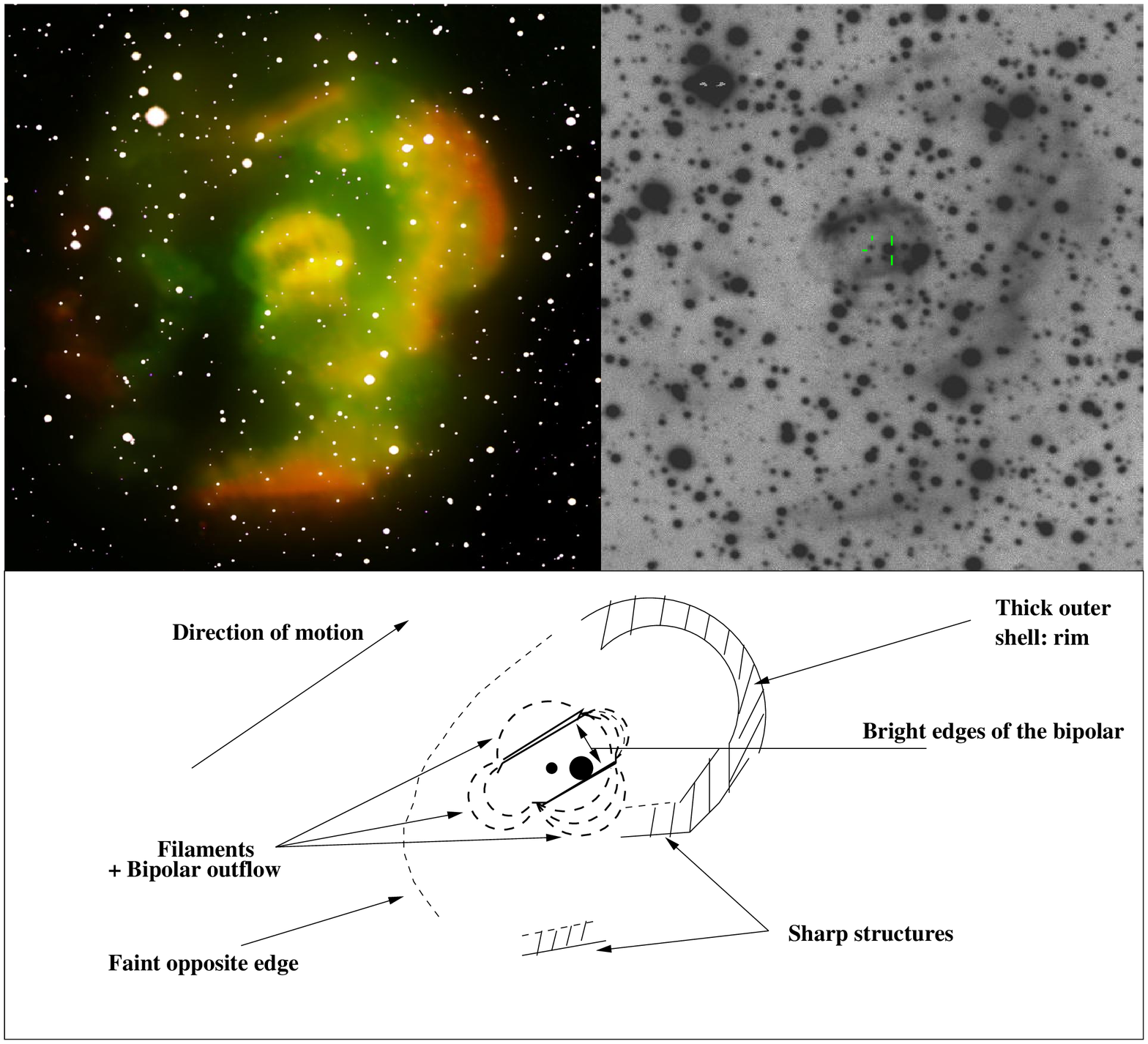}
\caption{An example of WZO1 type: The ``Ear Nebula'' IPHAS PN. North on the top and East on the left.}\label{ear}
\end{center}
\end{figure*}

\begin{figure*}[h]
\begin{center}
\includegraphics[scale=0.52, angle=0]{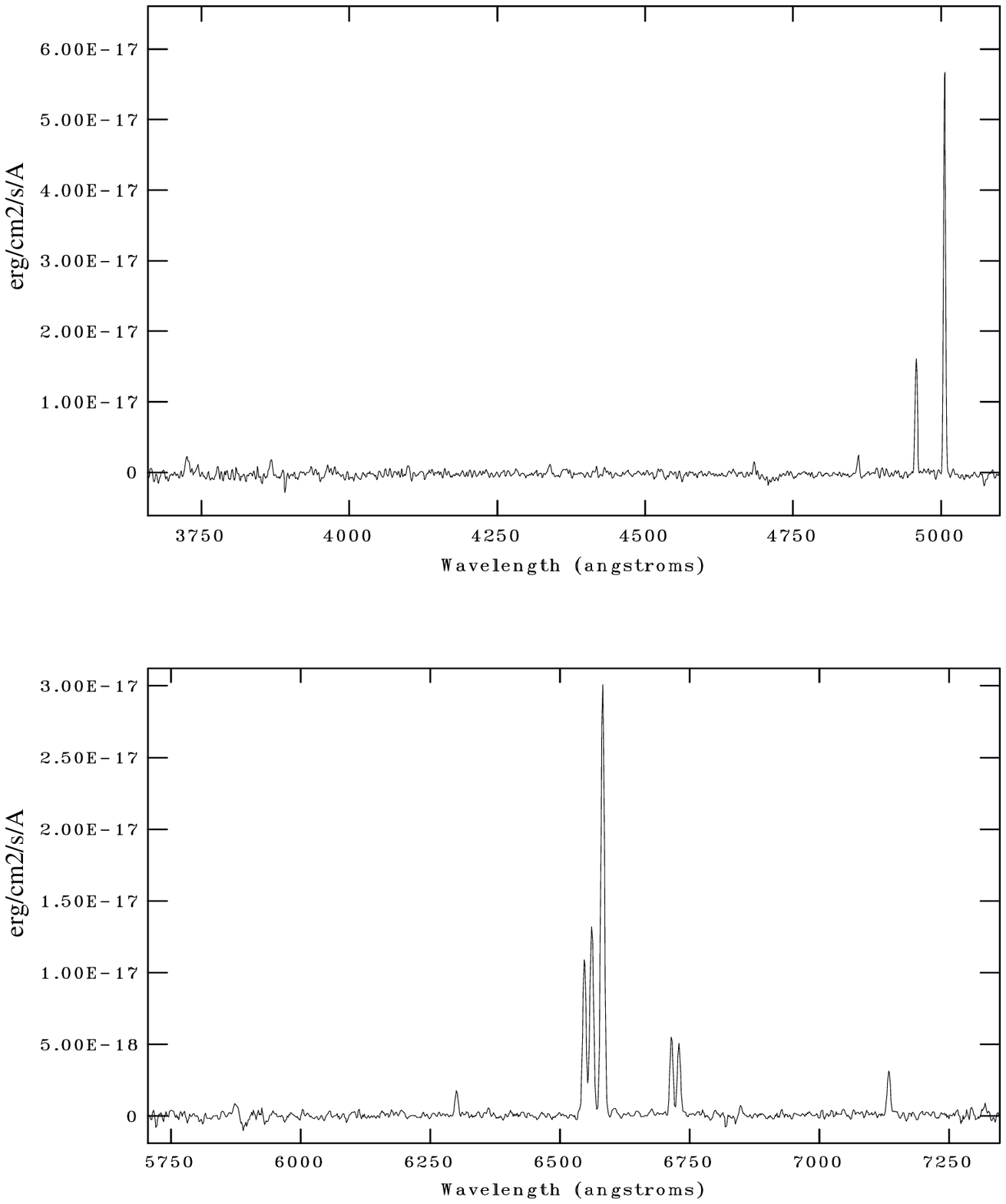}
\caption{WHT spectra of the ``Ear Nebula'' using the R300B and R158R gratings. This nebula, for which we show some of the ``strongest'' emission lines useful for an identification, presents a clear [NII] over-intensity and it has been confirmed as true PN using the revised diagnostic diagram from \citet{Riesgo06} (particularly the log [H$\alpha$/[SII]] {\it vs} log [H$\alpha$/[NII]] diagram).}\label{ear_spec}
\end{center}
\end{figure*}

The first group of PNe/ISM concerns those where the main PN is still
unaffected and which may display a distant bow shock. In our area of study
(18h-20h), the majority of candidates answer the first condition, but none
show the outer bow shock. Outside this area, the nicknamed ``Ear Nebula'' or
IPHASX J205013.7+465518 with a 6 arcmin size may be coincident with a WZO1
description as this object is a confirmed bipolar PN (Fig. \ref{ear_spec})
surrounded by a shell which may be an AGB remnant shell or would indicate a
multiple shell nebula (Fig. \ref{ear}).

\subsection{WZO2 type}

\begin{figure*}[h]
\begin{flushleft}
\includegraphics[scale=0.4, angle=0]{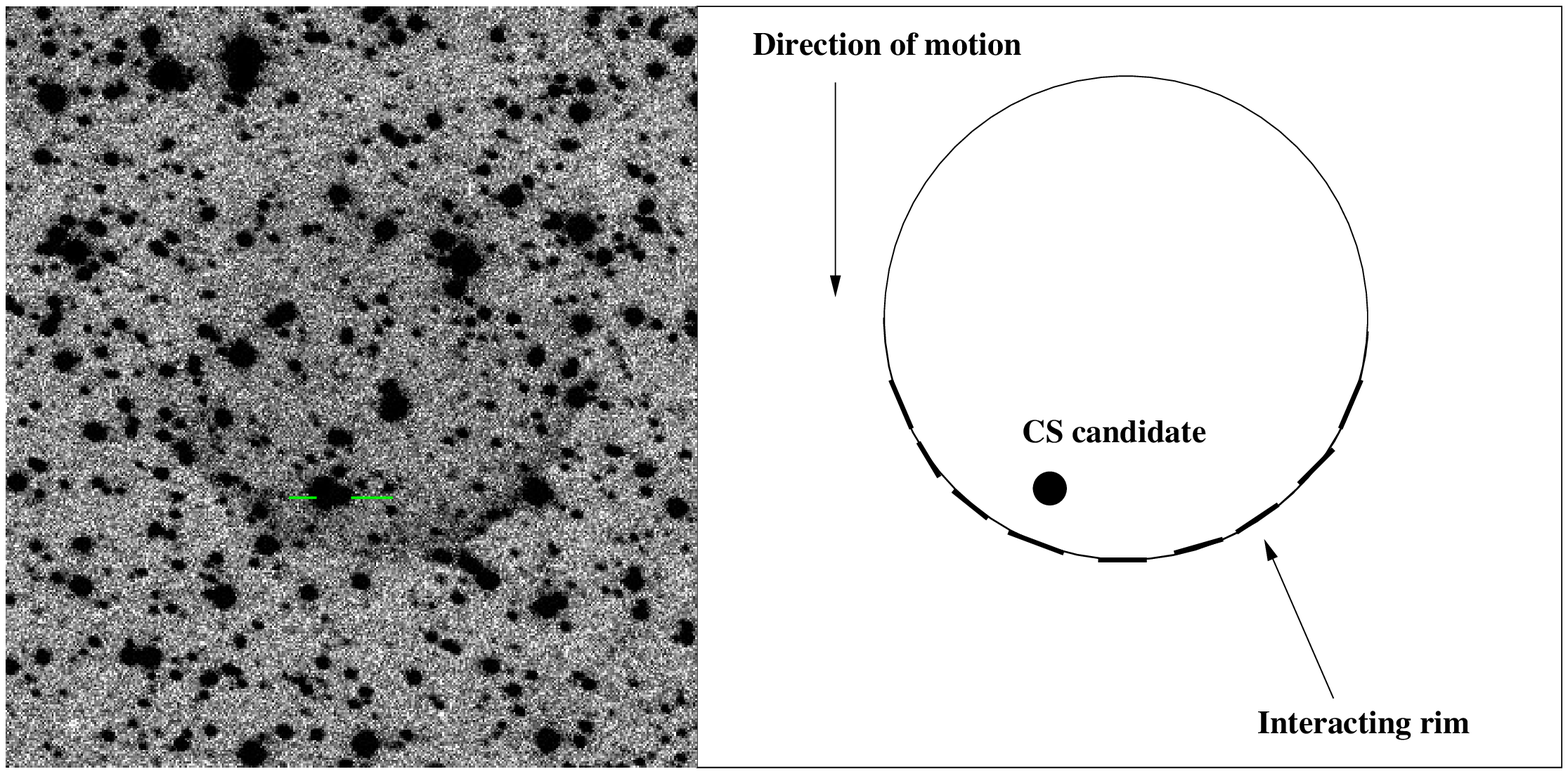}
\includegraphics[scale=0.4, angle=0]{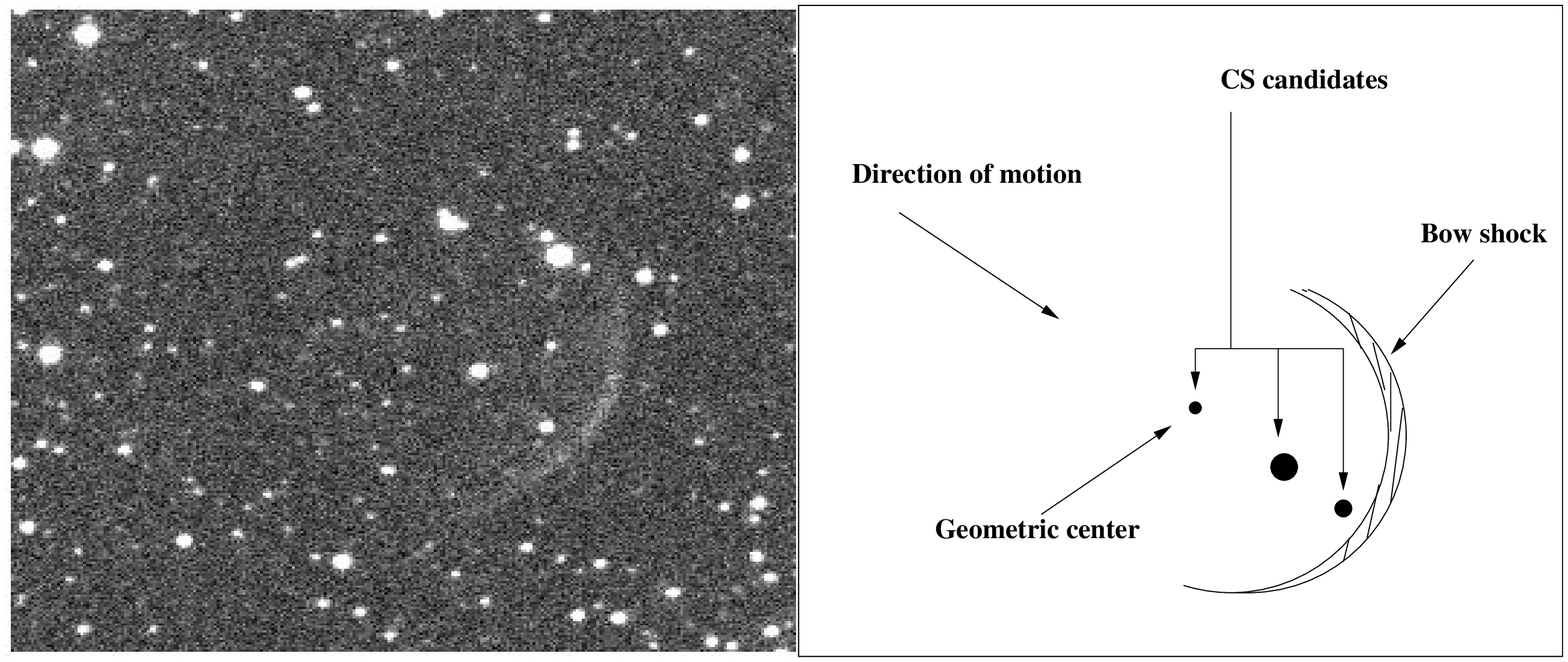}
\includegraphics[scale=0.4, angle=0]{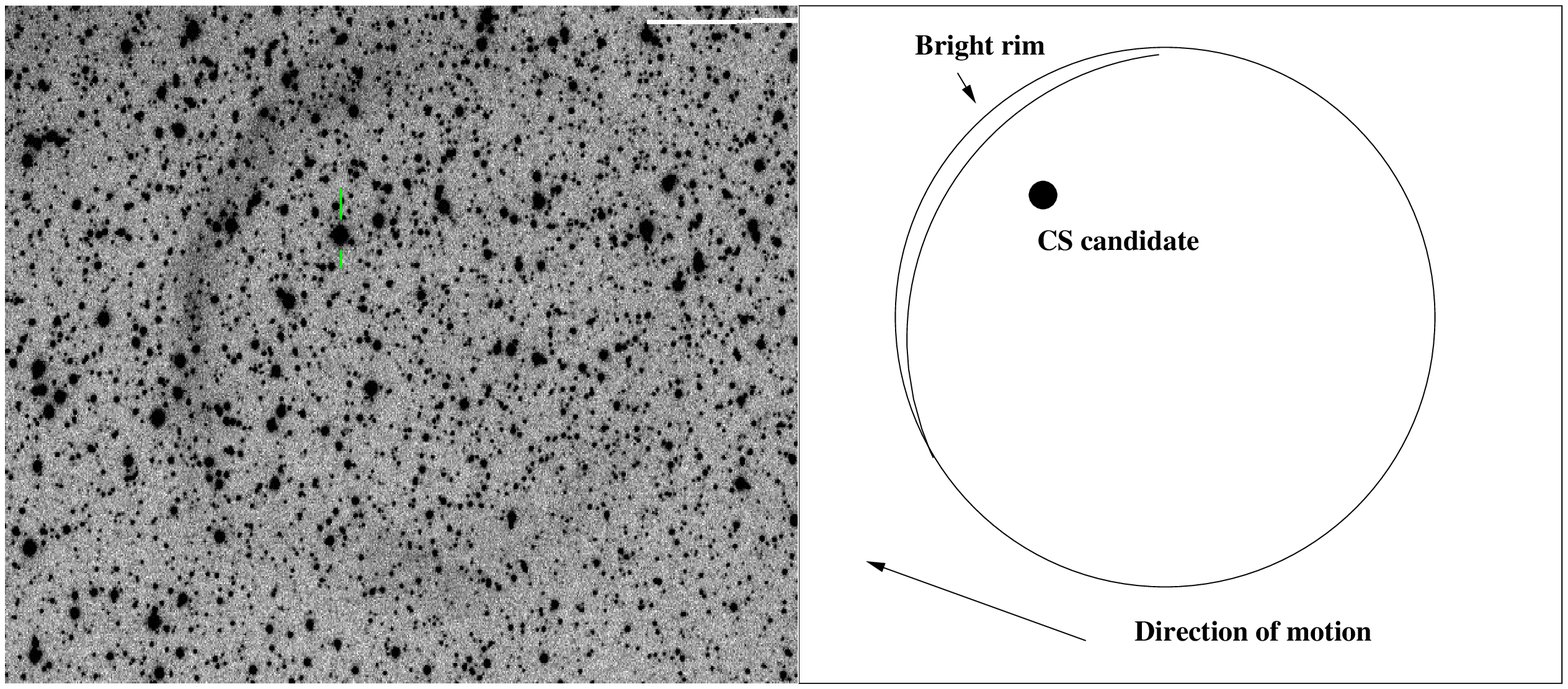}
\caption{Examples of WZO2 types. Top: Size=1.2 arcmin and SB=3.4e$^{-17}$ erg cm$^{-2}$ s$^{-1}$ arcsec$^{-2}$. Middle: Size= 8.5 arcmin and SB=1.1e$^{-16}$ erg cm$^{-2}$ s$^{-1}$ arcsec$^{-2}$, Bottom: 4.3 arcmin and SB=2.7e$^{-16}$ erg cm$^{-2}$ s$^{-1}$ arcsec$^{-2}$. North on the top and East on the left.}\label{wzo2}
\end{flushleft}
\end{figure*}

This category concerns PNe showing a bright rim in the direction of
motion. This is the most common feature found in our sample and 17 objects out
of 21 fall under this classification. Fig. \ref{wzo2} presents 3 examples with
different angular sizes, although they all display a diameter on the order of
a few arcmin (we considered the assumed full extent of the round nebulae). We
point out in Fig. \ref{wzo2}-Top the difficulty to determine the true
direction of motion regarding the CS position and off-axis bow shock. Such a
geometry could be explained by an ISM gradient from high on the left to low on
the right. We also notice a particularly low observed surface brightness (SB) which
may explain previous non detections.

\subsection{WZO3 type}

\begin{figure*}[h]
\begin{flushleft}
\includegraphics[scale=0.4, angle=0]{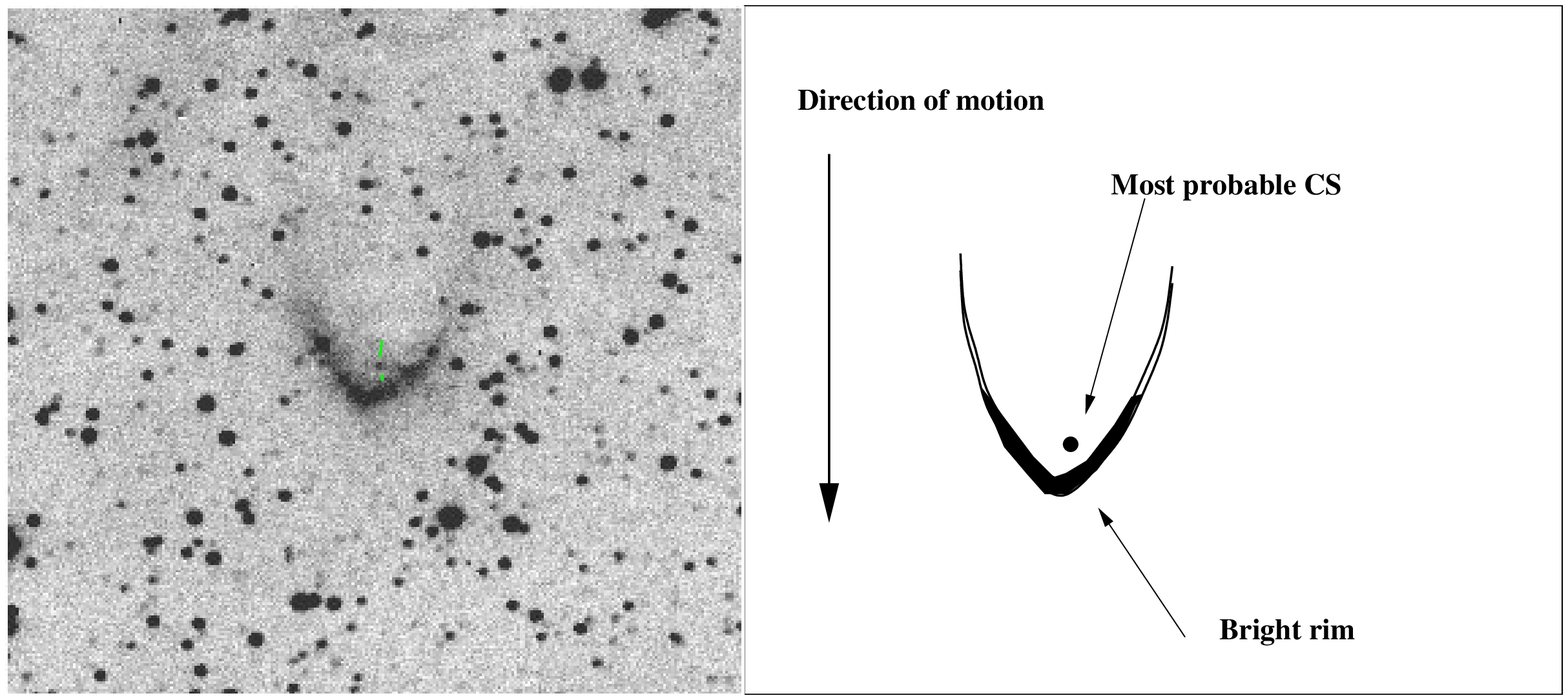}
\caption{WZO3 type of ISM interaction in a IPHAS candidate PNe. North on the top and East on the left.}\label{wzo3}
\end{flushleft}
\end{figure*}

This type is exemplified by PNe whose geometric centres are shifted away from
the central star (CS): both are no longer coincident. An example, is the
ancient PN Sh 2-188 around which IPHAS has uncovered an extended structure
\citep{Wareing2006}. We identified 3 candidate PNe coincident with this
description. The most probing WZO3 type in our sample is presented in figure
\ref{wzo3} and corresponds, according to hydrodynamical models, to a PN with a
CS velocity of about 100 km/s.

\subsection{WZO4 type}

The WZO4 corresponds to the most difficult types of PN to be detected: the CS
has left the vicinity of the now totally disrupted PN, leaving an amorphous
structure. The challenge does not lie in the detection ability (it enters in
the IPHAS range of detection) but more in the selection of the objects as
possible PNe due to the total lack of symmetry or axi-symmetry. This type of
interaction is also discussed in more detail by Wareing et al. in these
proceedings. 

\begin{figure*}[h]
\begin{center}
\includegraphics[height=4.9cm, angle=0]{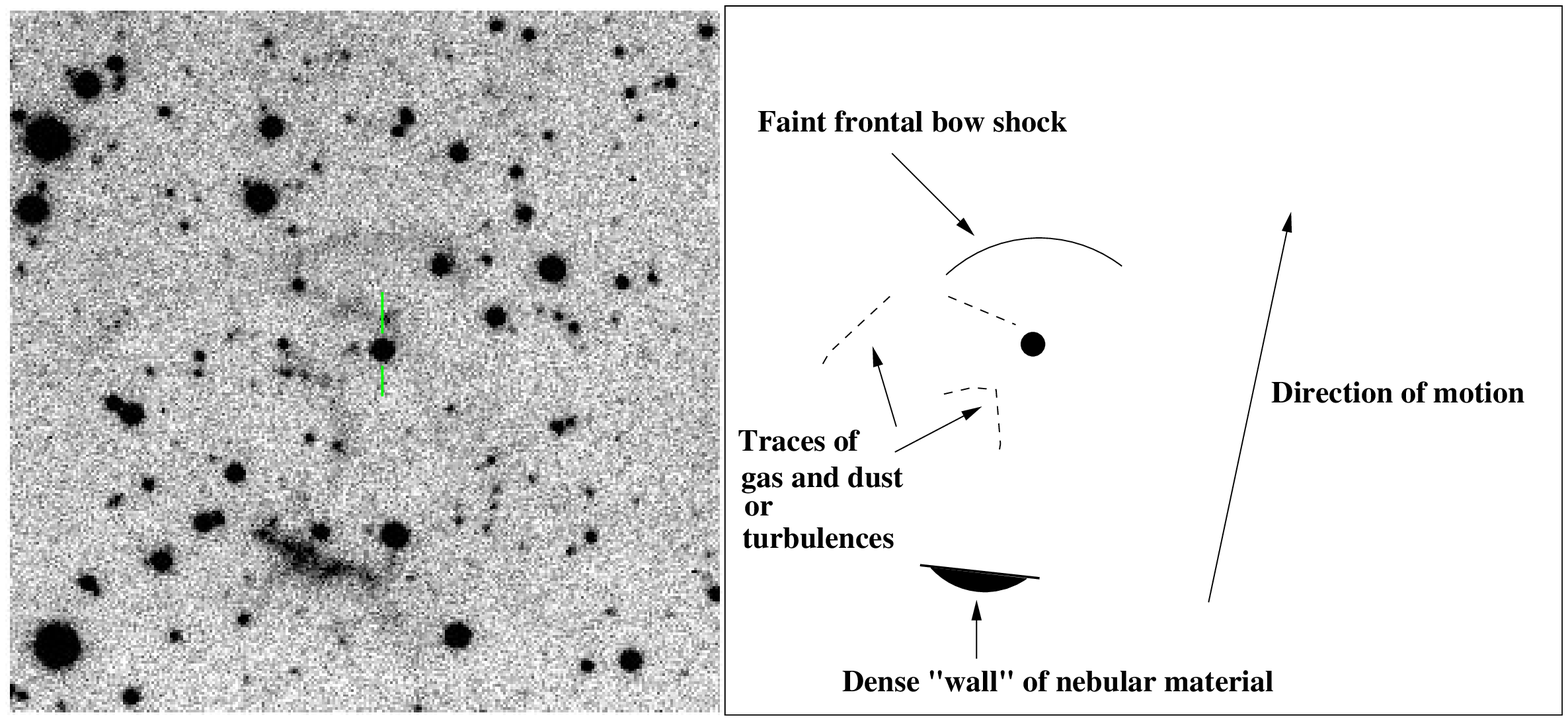}
\includegraphics[height=5.5cm, angle=0]{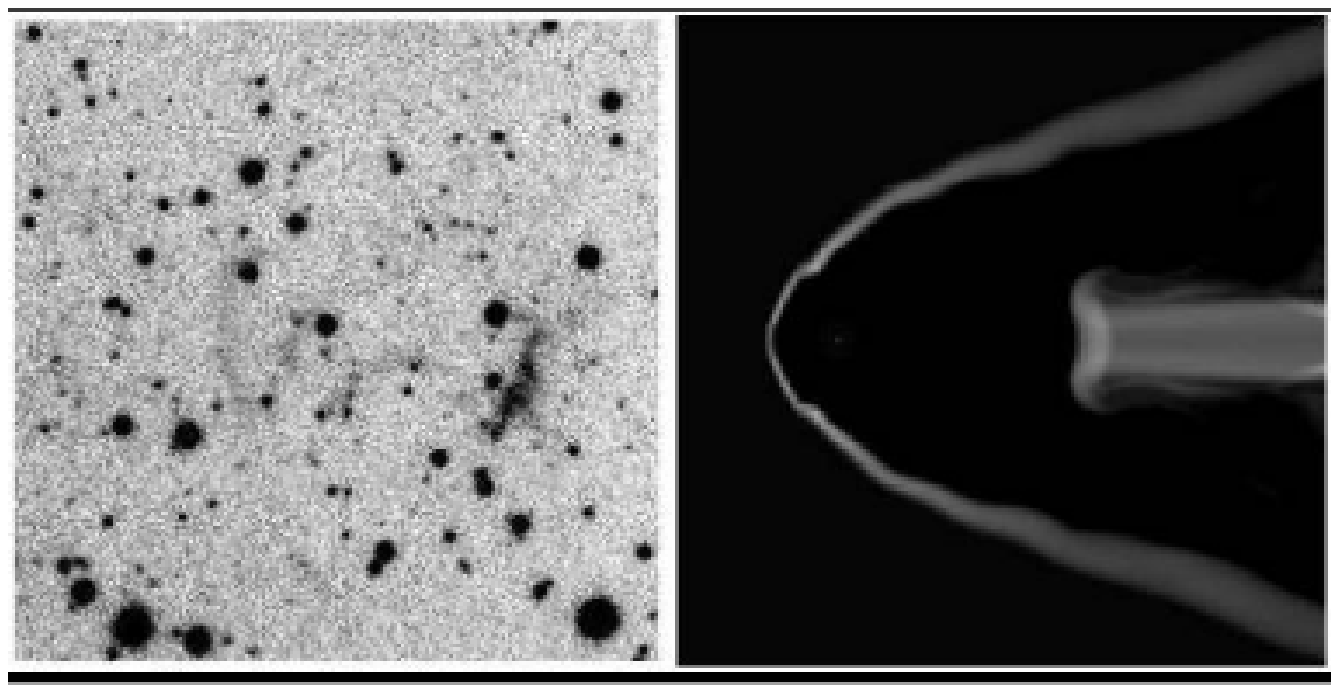}
\caption{ A possible example of WZO4 ISM interaction in one IPHAS PN candidate (top: North on the top and East on the left) with the corresponding hydrodynamical model (bottom) [reproduction of figure 5(d) from \citet{Wareing07}].}\label{wzo4}
\end{center}
\end{figure*}

We identified 1 candidate PN which could fit the given
description. Fig. \ref{wzo4} presents the selected candidate in the top
panel. We suggest the the nebular material has been moved from the front to
the rear leaving a remnant ``wall of material''. We also notice that some
features may be linked to turbulence effects. The comparison with the
hydrodynamical model (bottom panel) seems to support this
hypothesis. Nevertheless a spectroscopic confirmation of the nebula's nature
will be needed. The model implies a velocity relative to the ISM of 100 Km/s
and an evolution in the post-AGB phase of 10 000 years.

\subsection{Distribution of the candidates}

\begin{figure*}[h]
\begin{center}
\includegraphics[scale=0.8, angle=0]{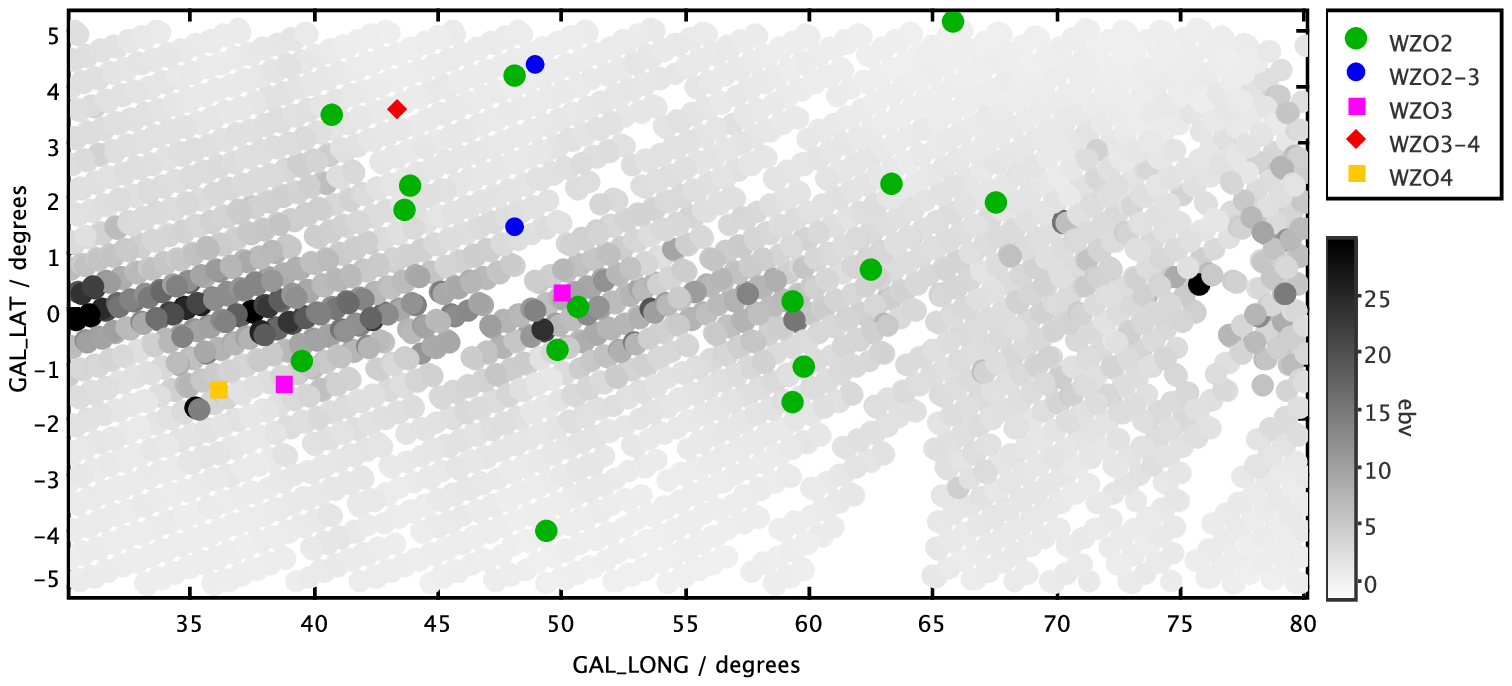}
\includegraphics[scale=0.8, angle=0]{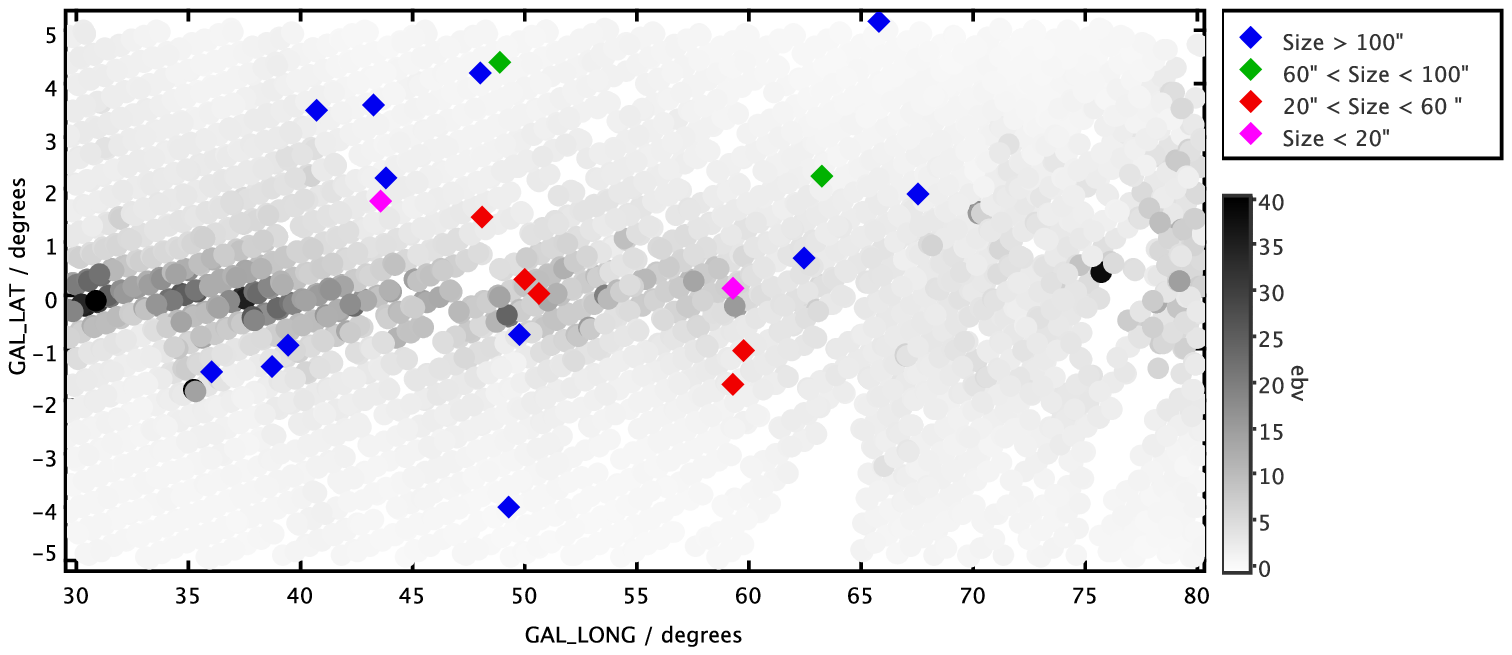}
\caption{ Galactic distribution of the candidate PNe/ISM according to their stage of interaction and their size.}\label{distrism}
\end{center}
\end{figure*}

Fig. \ref{distrism}-top shows that the majority of the WZO2 nebulae types are
located in zones of relatively low ISM density (compared to the Galactic
Centre). The low stress exerted on the nebulae may explain why they still
keep their quasi circular shape. The ISM is more dense in the Galactic Plane
than in the zone towards the anti-centre or the zone above a height of 100 pc
(from observation of neutral hydrogen gas, \citet{Dickey1990}). We therefore
expected a greater influence of the interaction process in this area. Indeed,
we observed that the most advanced stages of interaction, namely WZO3 and
WZO4, are detected in areas of high ISM density, where PN are more likely to
be affected by such densities. 

The size distribution, Fig. \ref{distrism}-bottom, indicates that although
most of the detected candidate PNe are large\footnote{The sizes here are defined in terms of angular sizes, so the physical correspondence will depend on the distance.}, i.e with a size greater than 100
arcsec, or of medium size i.e. between 20 and 100 arcsec, small nebulae also
show signs of interaction. This confirms that the ISM interaction process does
not ``a priori'' only imply ``old'' nebulae. We also observe that large
objects mainly lie at higher latitudes than smaller nebulae but it is also
interesting to notice that we detect large objects in zones of high
extinction; large PNe seem to survive at relatively low latitudes. They would
undergo strong alteration by the ISM and would display more advanced stages of
interaction. Those disruptions tend to affect them more than smaller size
nebulae at the same latitude range.


\begin{figure*}[h]
\begin{center}
\includegraphics[scale=0.35, angle=0]{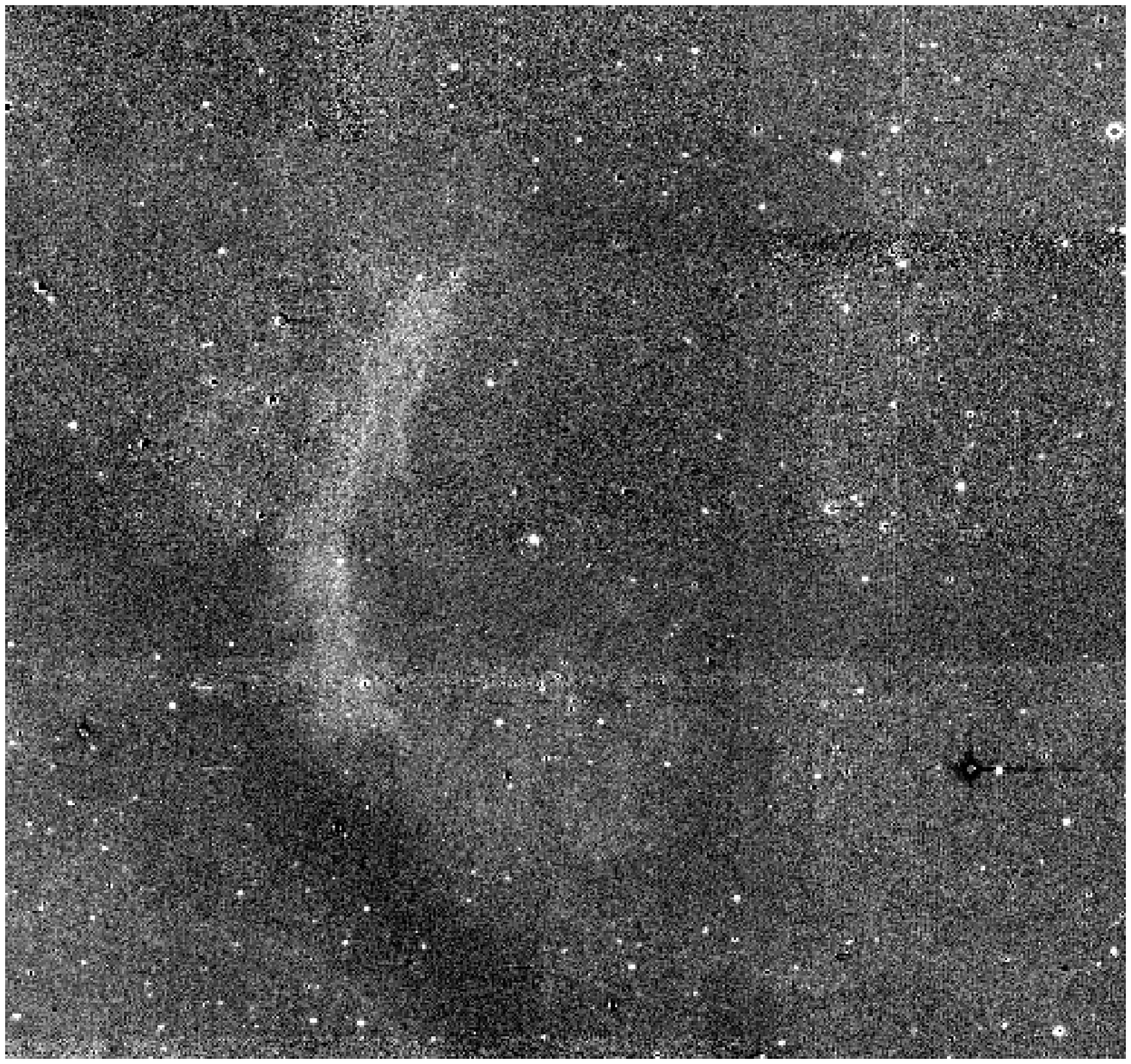}
\includegraphics[scale=0.38, angle=0]{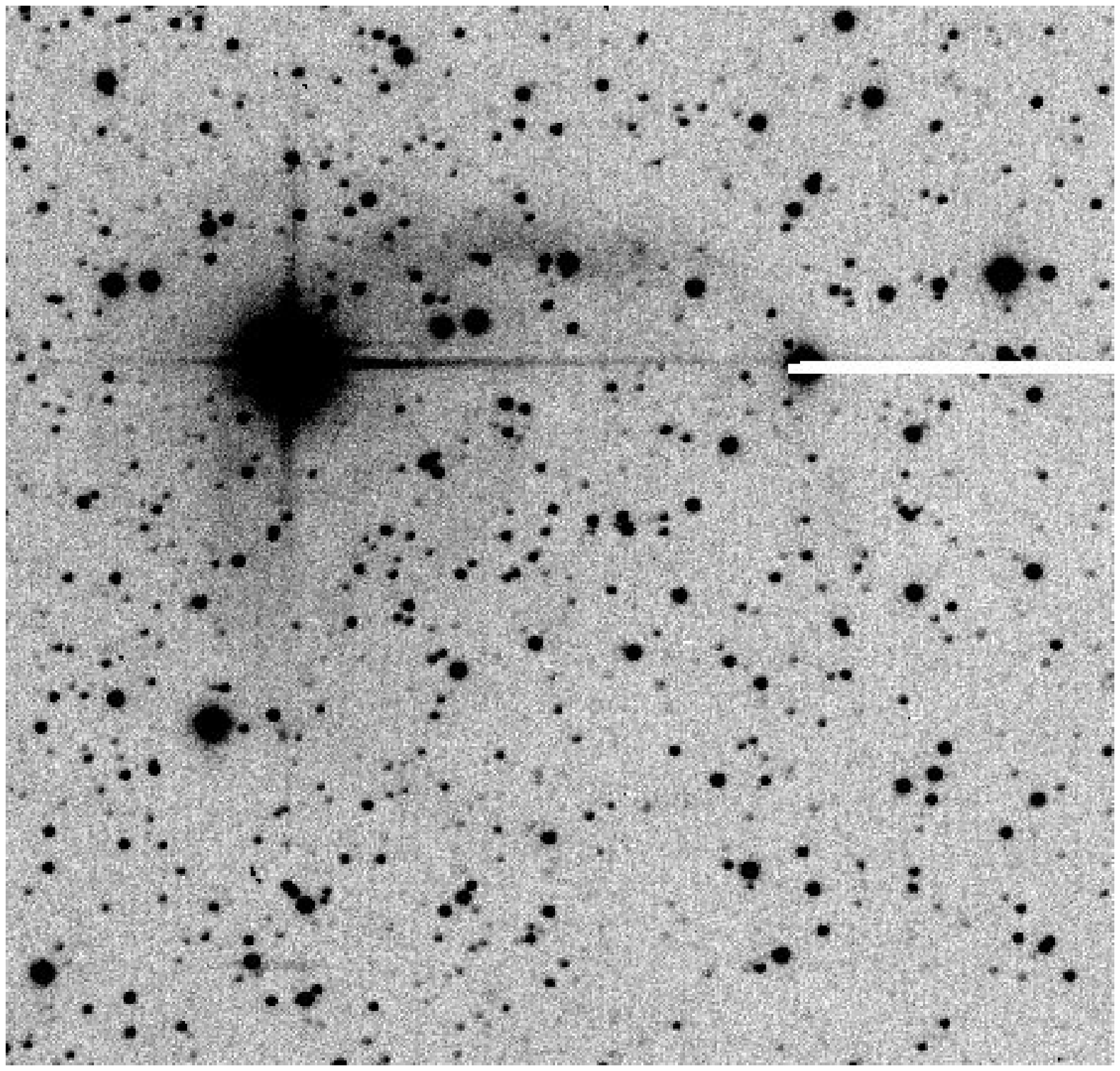}
\caption{Example of candidates with sizes greater than 100 arcsec (respectively 7.7 and 2.9 arcmin) and located at b=$\pm$1 deg. These objects present a WZO2 stage of interaction and only their (very) faint interacting rim are seen.}\label{large}
\end{center}
\end{figure*}

\section{Conclusion and Perspectives}

In the first  fully analysed area of the Galactic plane, RA=18h to RA=20h, the
new H$\alpha$ photometric survey IPHAS appears to be an excellent tool to
study PNe interacting with the ISM. Indeed the survey contributes to the
detection of nebulae so far hidden mainly due to their faintness.  Thus, 21
objects have been identified as possible planetary nebulae interacting with
the ISM. They show diverse sizes (although the majority display a diameter
greater than 100 arcsec) and morphologies corresponding to the four different
cases of interaction commonly defined going from the unaffected to the totally
disrupted nebula. The most common stage is the WZO2 corresponding to nebulae
showing a brightening of their rim in the direction of motion. This is
coincident with the observations made by Wareing and al (in these proceedings)
crossing different H$\alpha$ surveys.  We were also able to reach those
targets at low latitudes and found that some could survive in those
environments although they would be strongly affected by the ISM. The total
lack of PNe/ISM at the highest point of ISM density (b=$\pm$0.5 deg and 30 deg
$<$l $<$ 50 deg) can either be due to the limitation of IPHAS or because they
have been totally destroyed by the effects of ISM interaction.\\ The next
logical step is the spectroscopic identification of these sources, their
central star study and physical size determination. The low surface brightness implies the use of particular
means such as integral field spectroscopy to be able to retrieve the maximum
information. Therefore a new programme of IPHAS PN candidate follow-up
spectroscopy led by Q. Parker, A. Zijlstra and R. Corradi is now underway.\\




\end{document}